\documentclass[twocolumn,showpacs,unsortedaddress,superscriptaddress,showkeys,nofootinbib,preprintnumbers,letterpaper,aps,prd]{revtex4-2}
\usepackage{bm}
\usepackage{upgreek}
\usepackage{acronym}
\usepackage{amsmath}
\usepackage{amssymb}
\usepackage{natbib}
\usepackage{orcidlink}

\usepackage{here}
\usepackage{multirow}
\usepackage{graphicx}
\usepackage{enumitem}
\usepackage{lineno}

\usepackage{siunitx}
\usepackage{xcolor}
\usepackage{url}
\usepackage{comment}
\usepackage{xspace}
\usepackage[normalem]{ulem}

\usepackage{hyperref} 

\newlist{s5s6list}{itemize}{1}
\setlist[s5s6list]{label=\textbf{S5:}}
\newcommand\itema{\item[\textbf{S6:}]}

\newlist{catlist}{itemize}{1}
\setlist[catlist]{label=\textbf{CAT1:}}
\newcommand\itemcatb{\item[\textbf{CAT2:}]}
\newcommand\itemcatc{\item[\textbf{CAT3:}]}

\newcommand{\ihope}{\texttt{ihope}}

\newcommand{\figref}[1]{Fig.~\ref{#1}}
\newcommand{\secref}[1]{Sec.~\ref{#1}}
\newcommand{\tabref}[1]{Tab.~\ref{#1}}

\newcommand{\Msun}{\mathrm{M}_{\odot}}	
\DeclareSIUnit\Msun{M_{\odot}}	
\DeclareSIUnit\annum{a}		
\DeclareSIUnit\AU{AU}		
\DeclareSIUnit\parsec{pc}	

\begin{document}



\title{GW070605: An Undisclosed Binary Neutron Star Hardware Injection in LIGO's Fifth Science Run}

\author{Heather Fong \orcidlink{0000-0003-2722-6918}}
\thanks{\href{https://www.youtube.com/watch?v=dQw4w9WgXcQ}{heather@phas.ubc.ca}}
\affiliation{University of British Columbia, Vancouver Campus, 2329 West Mall, Vancouver, British Columbia V6T 1Z4, Canada}
\affiliation{RESCEU, The University of Tokyo, Tokyo, 113-0033, Japan}
\author{Kipp Cannon \orcidlink{0000-0003-4068-6572}}
\affiliation{RESCEU, The University of Tokyo, Tokyo, 113-0033, Japan}
\affiliation{Graduate School of Science, The University of Tokyo, Tokyo 113-0033, Japan}
\author{Chi-Wai Chan}
\affiliation{RESCEU, The University of Tokyo, Tokyo, 113-0033, Japan}
\affiliation{Graduate School of Science, The University of Tokyo, Tokyo 113-0033, Japan}
\author{Richard N. George \orcidlink{0000-0002-7797-7683}}
\affiliation{Center for Gravitational Physics, University of Texas at Austin, Austin, TX 78712, USA}
\author{Alvin K. Y. Li \orcidlink{0000-0001-6728-6523}}
\affiliation{RESCEU, The University of Tokyo, Tokyo, 113-0033, Japan}
\affiliation{Graduate School of Science, The University of Tokyo, Tokyo 113-0033, Japan}
\author{Soichiro Kuwahara}
\affiliation{RESCEU, The University of Tokyo, Tokyo, 113-0033, Japan}
\affiliation{Graduate School of Science, The University of Tokyo, Tokyo 113-0033, Japan}
\author{Hiroaki Ohta}
\affiliation{RESCEU, The University of Tokyo, Tokyo, 113-0033, Japan}
\affiliation{Graduate School of Science, The University of Tokyo, Tokyo 113-0033, Japan}
\author{Minori Shikauchi}
\affiliation{RESCEU, The University of Tokyo, Tokyo, 113-0033, Japan}
\affiliation{Graduate School of Science, The University of Tokyo, Tokyo 113-0033, Japan}
\author{Leo Tsukada \orcidlink{0000-0003-0596-5648}}
\affiliation{Department of Physics and Astronomy, University of Nevada, Las Vegas, 4505 South Maryland Parkway, Las Vegas, NV 89154, USA}
\affiliation{Nevada Center for Astrophysics, University of Nevada, Las Vegas, NV 89154, USA}
\author{Takuya Tsutsui}
\affiliation{RESCEU, The University of Tokyo, Tokyo, 113-0033, Japan}
\affiliation{Graduate School of Science, The University of Tokyo, Tokyo 113-0033, Japan}

\begin{abstract}
The authors wished to document the sensitivity improvement that has been contributed to the GW detection rate by detection algorithm research and development efforts, and set about re-analyzing S5 and S6 to determine the sensitive time-volumes of a modern pipeline and compare them to that of analysis algorithms of the day.  To our surprise, this effort led to the discovery of GW070605, what at first appeared to be a previously unreported high significance binary neutron star merger at a time when only the Livingston detector (L1) was operating---data that could not have been analyzed and a signal that could not have been discovered previously because the algorithms of the day required coincidence between two or more detectors.  GW070605's end time occurs in LIGO's L1 detector at 2007-06-05 18:37:02 UTC, and is estimated to be a merger with component masses of \qty{1.82}{\Msun} and \qty{1.24}{\Msun}.  The GstLAL detection algorithm estimates that noise processes produce false positives at least as significant as GW070605 at a rate of $8.6\times10^{-10}$ per year.  Disappointingly, subsequent investigations revealed the presence of a previously undocumented hardware injection in the L1 detector's Y arm end test mass' excitation channel, whose time and properties match that of GW070605.  The injection does not appear in the Gravitational Wave Open Science Center list of hardware injections.  We determined that while there is no sensitivity improvement between GstLAL and previous algorithms at the null-result threshold, there is marked improvement at above null-result thresholds; specifically, an approximately 55-times detection rate increase from initial-era algorithms at a \ac{FAR} threshold of 1 per 7000 years.  
\end{abstract}


\maketitle
\acrodef{FAP}{false-alarm probability}
\acrodef{FAR}{false-alarm rate}
\acrodef{GW}{gravitational wave}
\acrodef{GWOSC}{Gravitational Wave Open Science Center}
\acrodef{LIGO}{Laser Interferometer Gravitational-wave Observatory}
\acrodef{RV}{random variable}
\acrodef{SNR}{signal-to-noise ratio}
\acrodef{BNS}{binary neutron star}
\acrodef{BBH}{binary black hole}
\acrodef{NSBH}{neutron star black hole}
\acrodef{IMBH}{intermediate mass black hole}
\acrodef{ETM}{end test mass}
\acrodefplural{ETM}[ETMs]{end test masses}
\acrodef{CBC}{compact binary coalescence}
\acrodef{S5}{fifth science run}
\acrodef{S6}{sixth science run}


\section{Introduction}\label{sec:intro}

Like a graduate student chasing a free lunch, the world in 2005 was moving as fast as it knew how. Yet, two decades later, retrospection has a way of slowing down the past. Back then, the Internet was quainter, quieter; it was a virtual wild west you explored on a clunky, beige steed of a computer that, by today's standards, would be put out to pasture the same day it came home. If you wanted to know something about something, you asked Yahoo, Google, or Alta Vista, and then you waited... and waited. Knowledge inched in scanline by scanline, but it still left analogue learning in the dust. It was the age of Web 2.0, where people flipped phones, burned CDs, surfed the 'net, and still left tweeting to the birds. 

One thing that hasn't changed, however, is the universe's tendency to hold fast to its secrets, and our tendency to try coaxing them loose. 

In November 2005, the first-generation \ac{LIGO} \cite{LIGOScientific:2007fwp} began its \ac{S5} in search of \acp{GW}, which had not yet been directly detected. A two-year run for the Hanford and Livingston LIGO detectors, they were later joined by the Virgo detector \cite{VIRGO:2012dcp} for a joint observation before the three detectors concluded their science runs in October 2007 \cite{Abadie_2010,Abadie_2011}. The verdict? No \ac{GW} signals. But the teams were undeterred, and two years later, the detectors were back online to kick off \ac{LIGO}'s \ac{S6} and Virgo's 2nd science run, with better hardware, better technology, better sensitivity \cite{Abadie_2012}. Observations concluded in October 2010, and disappointingly, analyses published at the time reported null results. But as we all know now, the search for \acp{GW} was far from over, and history was still in the making \cite{gw150914,gw151226,O1paper,gw170814,gw170817,gw170104,gw170608,GWTC_1,GWTC_2,gw190412,gw190425,gw190521,gw190814,gw200105_gw200115,GWTC_3,gw230529}.

Sixteen years after the end of \ac{S6}, we're now in the thick of the advanced era of ground-based \ac{GW} detectors \cite{KAGRA:2013rdx,LIGOScientific:2014pky,VIRGO:2014yos,KAGRAoverview}. Not only the detectors themselves have improved, but the detection algorithms---the thousands of lines of code that combs through every point of data to tease out signals from the noise---have as well. This brings us to the main question we aim to answer: how do the sensitivities of initial-era detection pipelines compare to that of a modern one? In this paper, we describe and present the results of our study, which comprises a full analysis of \ac{S5} and \ac{S6} strain data using the GstLAL detection pipeline \cite{Messick_2017,sachdev2019gstlalsearchanalysismethods,Hanna_2020,cannon2020gstlalsoftwareframeworkgravitational}, a matched filter-based search algorithm developed for the advanced era of \ac{GW} searches for \acp{CBC}. But no good story is without its surprises, and ours takes an unexpected turn when GstLAL detects a highly significant signal that, at first, appears to originate from a binary neutron star merger. However, a deeper investigation reveals a second twist, that the signal's true origin is not astrophysical, but in fact, a hardware injection.

The paper is laid out as follows. In Section~\ref{sec:data}, we give a description of the data used in our analysis. The detection pipelines, past and present, are described in Section~\ref{sec:search}, with additional details on the search parameters specifically relevant to the analysis. Our results are presented in Section~\ref{sec:results}, with a discussion of the most significant event, GW070605, and a comparison of the sensitive spacetime volumes between the GstLAL and initial-era detection pipelines. Section~\ref{sec:conclusion} concludes the main study with a summary of the key results. We also include a brief, bonus study in Appendix~\ref{app:virgo}, in which we determine whether or not the inclusion of Virgo in the \ac{GW} detector network affects the sensitivity of the GstLAL detection pipeline.


\section{Data}\label{sec:data}

Using the public data provided by the \ac{GWOSC}, we analyzed strain data taken during \ac{S5} and \ac{S6}, the fifth and sixth science runs of initial \ac{LIGO}, respectively. The durations of each science run are as follows:
\begin{s5s6list}
\item November 4, 2005 16:00:00 UTC to October 1, 2007 00:00:00 UTC
\itema July 7, 2009 21:00:00 UTC to October 20, 2010 15:00:00 UTC
\end{s5s6list}

Data during times spanned by CAT1 vetoes---where the data are classified as unusable for astrophysical searches---and around the times of software and hardware injections reported in \ac{S5} \cite{Abadie_2010} and \ac{S6} \cite{Abadie_2012} were omitted, including blind injections. In addition, internal \ac{LIGO} records were checked for hardware injections that had not been recorded on \ac{GWOSC} or in other public \ac{LIGO} Scientific Collaboration documents.  At the time of this study; we confirmed all documented injections had already been identified in \ac{GWOSC}'s segment lists.

Our analysis is limited to data collected by the \ac{LIGO} Hanford and Livingston detectors, known also as H1 and L1. Although the second Hanford detector, H2, was also in observing mode during \ac{S5}, the GstLAL detection pipeline's noise model assumes that different detectors experience statistically independent noise processes. Since H2 was co-located with H1, with the two sharing a vacuum envelope and their beam splitters and input test masses being in close physical proximity, their noises are not independent; therefore, H2, the less sensitive of the two Hanford detectors, is excluded from the analysis. Data collected by the GEO600 and Virgo detectors were also omitted\footnote{Brief stretches of Virgo strain data from its science runs VSR1 and VSR2 are used in the study described in Appendix~\ref{app:virgo}.} as their strain data are unavailable on \ac{GWOSC}.


\section{Search}\label{sec:search}


\subsection{Pipelines}\label{sec:pipelines}

Previous analyses of \ac{S5} and \ac{S6} utilized a matched filtering search pipeline (later known as \ihope), the details of which are described in \cite{Allen_2012,Babak_2013}.  To summarize, the pipeline's algorithm sets a \ac{SNR} threshold as a preliminary test to identify potentially interesting events known as \textbf{triggers}.  Each trigger has a binned frequency-domain \(\chi^{2}\) residual test statistic computed for it \cite{Allen_2012}, which is combined with the \ac{SNR} in an \textit{ad hoc} non-linear expression to yield a re-weighted \ac{SNR}, the sum-of-squares of which, across detectors, was used to rank triggers and differentiate signals from noise. The rate of false positives was estimated by performing time-shifted analyses and treating time-shifted coincident triggers as false alarms. The pipeline only analyzed triggers found in coincident time (\textit{i.e.}\@ when two or more instruments were online at the same time).  No attempt was made to identify signals during times when only one instrument was online due to the inability of the algorithm to form a background model \cite{Babak_2013}.

In our analysis, we use the GstLAL detection pipeline \cite{Messick_2017,sachdev2019gstlalsearchanalysismethods,Hanna_2020,cannon2020gstlalsoftwareframeworkgravitational}, which has participated in all observing runs in the advanced eras of \ac{GW} searches \cite{GWTC_1,GWTC_2,2022KAGRA,GWTC_3}. Unlike the pipelines in the initial era, which used manually-tuned \textit{ad hoc} ranking statistics constructed exclusively from single-detector \ac{SNR} and \(\chi{^2}\) tests \cite{babak2012a}, GstLAL uses direct estimation of the signal and noise densities in a high-dimensional parameter space consisting of the \acp{SNR} and \(\chi^{2}\)-like residual tests from all detectors in the network together with the trigger phases, the time delays between a candidate's arrival in different detectors, all detector sensitivities to the waveform model at the time of the candidate whether the detectors saw it or not, and mean false-positive rates of the detectors at the time of the candidate.  Addtionally, an astrophysically motivated source population model weights candidates based on the intrinsic parameters of each candidate's template and the template density \cite{Fong2018elx}.

The algorithm trains the signal and noise distribution densities on the observed data, and together their ratio comprises a ranking statistic.  To the extent the training is successful, by the Neyman-Pearson lemma \cite{neyman}, the likelihood ratio provides a ranking statistic that extremizes detection efficiency at fixed \ac{FAR}.  Great care is taken to ensure the accuracy of the ranking statistic's denominator distribution density.  By assuming that distribution density is correct, and drawing artificial candidates from it using an importance-weighted sampling procedure, the GstLAL detection system can predict the distribution of ranking statistics to be observed from the noise process.  The ability to construct its noise model without the use of time-shifted coincidences enables the system to model the false-positive rate of single-detector candidates as easily as multi-detector candidates, and in this way, the system can search for signals at times when only one detector is available.

Previous detection algorithms treated all templates and all detectors as equal---\ac{SNR} is \ac{SNR} is \ac{SNR}.  This led to the problem that a given search's sensitivity was limited to the sensitivity of the least sensitive detector, because that detector was providing false alarms at a given \ac{SNR} threshold at the same rate as the others, but finding fewer signals.  For this reason, GEO600 was excluded from searches at the time.  That limitation was addressed by the ``{LVStat}'' ranking statistic, introduced to facilitate the inclusion of Virgo in searches for \acp{GW} from compact object collisions \cite{Abadie_2010}, as Virgo was significantly less sensitive than the \ac{LIGO} detectors.  GEO600, however, was still excluded from that search.  In contrast, because time-dependent per-template and per-detector horizon distances are included parametrically in the multi-dimensional density that forms the numerator of GstLAL's ranking statistic, there is no such limitation, and any detector with any sensitivity, indeed even a random number generator, may be included in a search.  A relatively insensitive detector adds to the computational cost but never diminishes the algorithm's sensitivity.  As stated above, we excluded GEO600 and Virgo from our reanalysis only because of the lack of public data.  We will show a small test with Virgo data to demonstrate the sensitivity improvement it provided to the network in Appendix~\ref{app:virgo}.

Because the information of which template produced a given candidate is included in the multi-dimensional distribution densities, not only can the template-specific sensitivities of the detectors be accounted for, the GstLAL algorithm also bases its assessment of candidates on an astrophysically motivated source population model.  This affords an opportunity to target a search at specific source types \cite{Fong2018elx}.  In this search, the source population model assumed binary systems whose component masses are distributed uniformly in their logarithms.

\begin{table}
\centering
\begin{tabular}{| c | c | c | c |}
\hline
{\bf Run} & {\bf Instruments} & \bf Previous/\ihope{} & \bf GstLAL \\
\hline
\multirow{5}{*}{S5} & H1H2L1 & 226 days & | \\
 & H1L1 & 35 days & 390 days \\
 & H2L1 & 22 days & | \\
 & H1 & | & 144 days \\
 & L1 & | & 64 days \\
 \cline{2-4}
& Total & 283 days & 598 days \\
\hline
\multirow{5}{*}{S6} & H1L1V1 & 51 days & | \\
 & H1L1 & 77 days & 138 days \\
 & H1V1 & 47 days & | \\
 & L1V1 & 29 days & | \\
 & H1 & | & 98 days \\
 & L1 & | & 80 days  \\
\cline{2-4}
& Total & 204 days & 316 days \\
\hline
\multicolumn{2}{| c |}{\bf S5+S6 Total} & \bf 487 days & \bf 914 days\\
\hline
\end{tabular}
\caption{Total time analyzed by each detection pipeline, where the times in the ``Previous/\ihope{}" column are taken from \cite{Abadie_2011} and \cite{Abadie_2012}, while the times in the GstLAL column correspond to the analysis detailed in \secref{sec:search}, using the data described in \secref{sec:data}. Since previous detection pipelines required events to be coincident, data where only one detector was observing was not analyzed; conversely, GstLAL does not have this requirement. Note that since H2 and V1 data were omitted from the GstLAL search, coincident times that include these detectors have been re-categorized accordingly--for example, H2L1 time is considered L1-only in the GstLAL column.}
\label{table:time}
\end{table}

The analysis times of the previously-published searches and GstLAL are given in \tabref{table:time}. The differences arise largely from the inclusion of single-detector times in the GstLAL-based search and the choice of veto categories:
\begin{catlist}
\setlength{\itemindent}{1.75em}
\item Data that are flagged to be severely compromised, to the point that they should not be used in searches for astrophysical signals.
\itemcatb Data that are flagged to contain a major instrumental problem at the time of collection.
\itemcatc Data that are flagged to contain a moderate concern at the time of collection.
\end{catlist}

In previous analyses, data flagged with CAT1, CAT2, and some CAT3 vetoes were excluded. Additionally, approximately 10\% of S6 data was used for manual tuning of the \textit{ad hoc} ranking statistic and data quality investigations and not searched for \ac{GW} signals \cite{Abadie_2012}. Previous experience has shown that \ac{GW} detection algorithms based on machine-learning multi-variate classifier systems like GstLAL exhibit higher detection efficiency at a given false-alarm rate threshold when vetoes are not applied to the data set.  For example, the search for bursts from cosmic strings in \ac{LIGO}'s fifth and sixth and Virgo's first through third science runs \cite{s56vsr123strings}, which made use of an early form of the same ranking statistic framework upon which GstLAL is built, exhibited this effect.  To improve its sensitivity, subsequent searches with that pipeline were conducted without the application of vetoes \cite{PhysRevD.97.102002}.  We speculate that the issue is that while vetoes remove some glitches from the data, they do not remove all of them.  Removing a subset of glitches, perhaps those that are most obvious, effectively removes training data, which diminishes the detection algorithm's ability to learn how to differentiate between glitches and signals.  Consequently, current glitch-removing procedures cause machine-learning based classifiers like GstLAL to \emph{increase} the rate of false positives, or diminish the detection efficiency at a given fixed false-positive rate.  Based on that past experience, we chose to only exclude data flagged as a CAT1 veto and data that contain hardware injections.


\subsection{Template bank}\label{sec:bank}

The template bank was constructed by generating waveforms with starting frequency $f_{\mathrm{min}} = \qty{30}{\hertz}$, using a stochastic placement algorithm \cite{Privitera_2014,Harry_2009} and distributing the templates densely enough such that no more than 10\% of detectable signals would be missed due to the discreteness of the template bank. Using the IMRPhenomD approximant as our waveform model \cite{Khan_2016}, which models the entire inspiral-merger-ringdown description in the frequency domain, the parameter space of our template bank includes non-precessing, dimensionless spin limits of
\begin{equation}
|\chi_{i}| \le \begin{cases}
	0.05, & m_{i} \le \qty{3}{\Msun}\\
	0.99, & m_{i} > \qty{3}{\Msun},
	\end{cases}
\label{eqn:spin}
\end{equation}
and, initially, component mass limits of $\qty{1}{\Msun} \le m_i \le \qty{100}{\Msun}$ (where $i=1,2$) and a total mass limit of $m_1 + m_2 \le \qty{100}{\Msun}$. However, under these mass limits, our initial search results revealed an excess of triggers that had been rung up by high mass ratio templates ($m_1/m_2 \gtrsim 10$). This was consistent with the search results reported in \cite{Abadie_2012}, which noted that templates with total mass $m_1+m_2 > \qty{25}{\Msun}$ were more susceptible to non-stationary noise in the data. As a result, we generated a new template bank that imposes the same mass limits as the template bank used in \cite{Abadie_2012}, where $m_1 + m_2 \le \qty{25}{\Msun}$. Spin limits were unchanged from \eqref{eqn:spin}. 


\subsection{Additional search parameters}\label{sec:otherparameters}

One method used by GstLAL to reduce the computational expense of matched-filtering is to decompose the template bank into a reduced set of orthonormal filters \cite{Messick_2017}. To prepare the templates for decomposition, GstLAL groups them into bins of similar time-frequency evolution using two parameters: the effective spin parameter $\chi_{\mathrm{eff}}$
\begin{equation}
\chi_{\mathrm{eff}} = \frac{m_1\chi_1 + m_2\chi_2}{m_1+m_2},
\end{equation}
and chirp mass $\mathcal{M}_c$,
\begin{equation}
\mathcal{M}_c = \frac{(m_1 m_2)^{3/5}}{(m_1+m_2)^{1/5}}.
\end{equation}
Each bin contains approximately 300 templates with similar $\chi_{\mathrm{eff}}$ and $\mathcal{M}_c$ values.

In addition, GstLAL's ranking statistic includes the template likelihood $P(t_k | \mathrm{signal}, \rho)$, which is the probability that a trigger is recovered by a template waveform $t_k$, given the trigger \ac{SNR}, $\rho$, and that the signal belongs to a population \cite{Fong2018elx}.  Intuitively, triggers that are the result of signals at high \ac{SNR} should be found distributed in intrinsic parameter space the way the sources are, while at low \ac{SNR} triggers that are the result of signals will be distributed uniformly across the templates.  Therefore, the template likelihood is informed by the distribution of templates within the bank and a signal population model, and depends parametrically on the \ac{SNR} of the candidate. For this search, we start with the following model:
\begin{equation}
p(\vec{\theta} | \mathrm{signal}, \rho)\mathrm{d}\vec{\theta} \propto \frac{1}{4 m_{1,i} m_{2,i}}\mathrm{d}\vec{\theta},
\end{equation}
where $\vec{\theta}=\{m_1, m_2, \chi_1, \chi_2\}$ are the template waveform's intrinsic parameters. This model has been used in advanced-era \ac{LIGO}-Virgo compact binary searches and is deliberately broad to minimize missed signals \cite{GWTC_2,GWTC_3}.

The ranking statistic's denominator density approximates false-positive rates and trigger parameter densities as independent of template within each template bank bin, whereas in reality sometimes a small number of the templates within that bin are responsible for a disproportionate fraction of the false positives.  Because the noise model draws artificial candidates from these densities, it will not model this phenomenon if it occurs, it will draw templates from the bin as if all are equally likely in the noise process.  To ensure this still yields the correct distribution of ranking statistics, this search used a slightly modified astrophysical population model for the numerator.  The template probabilities were averaged over the template bank bin, and all templates in each bin were assigned the same average probability mass for the bin.
\begin{equation}
P(t_k | \mathrm{signal}, \rho) = \frac{\sum_{i=1}^{N} P(t_i | \mathrm{signal}, \rho)}{N},
\label{eqn:p_tk}
\end{equation}
where $\{t_0, ..., t_N\}$ are templates that belong in the same bin.  This approximation diminishes the sensitivity of the ranking statistic, but ensures that the noise model predicts the correct relative frequency of ranking statistic values, without having to correctly model exactly which template within a given bin is responsible for the false positives.

\begin{figure}
\includegraphics[width=\columnwidth]{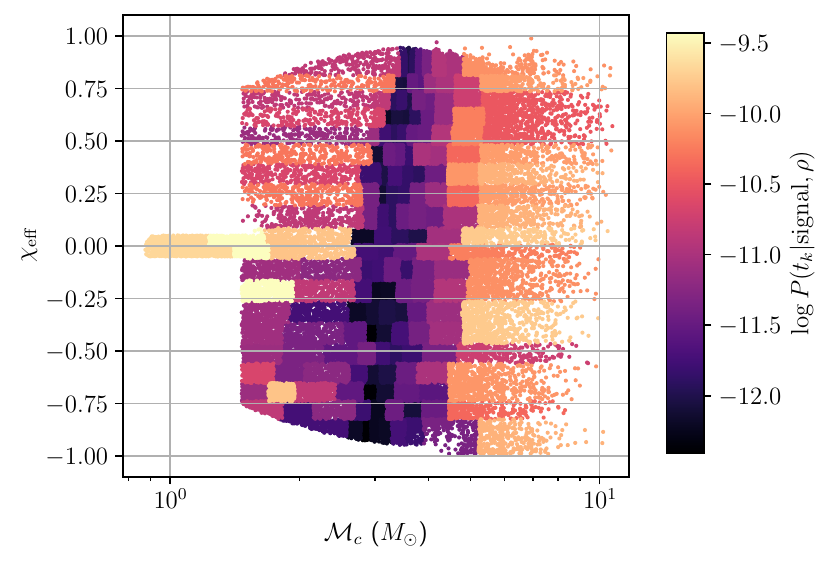}
\caption{Template waveforms plotted in terms of their chirp mass $\mathcal{M}_c$ and effective spin $\chi_\mathrm{eff}$. The total mass of the templates is limited to $m_1+m_1 \le 25 \Msun$ to reduce the number of triggers rung up by high mass ratio templates due to non-stationary noise \cite{Abadie_2012}. Colour denotes the template likelihood $P(t_k|\mathrm{signal},\rho)$ at $\rho=10$, and is approximated as a constant over each template bin. Each bin contains approximately 300 templates, which populate the bank via a stochastic placement algorithm \cite{Privitera_2014,Harry_2009}. The waveform approximant used is IMRPhenomD \cite{Khan_2016}.}
\label{fig:bank}
\end{figure}

\figref{fig:bank} shows the template likelihood distribution in the $\mathcal{M}_c-\chi_{\mathrm{eff}}$ parameter space at optimal \ac{SNR} $\rho=10$. Because templates in the same bin are assigned the same likelihood using \eqref{eqn:p_tk}, the coloured patches illustrate how the templates are grouped. Since all bins have approximately the same number of templates, the relative sizes of the patches also gives a sense of the template density in this parameter space.

\subsection{Identification of Signals}\label{sec:identification_of_signals}
For each candidate, the ranking statistic, $\ln\mathcal{L}$, is computed.  As described in \secref{sec:pipelines}, this is the natural logarithm of the ratio of the density of true signals to false positives in the parameter space describing the candidates.  To prevent $\ln\mathcal{L}$ from being dependent on the duration of the experiment (so that all candidates do not need to be relabelled if some small amount of additional data is analyzed), the numerator and denominator densities in that ratio are not normalized; for example, the densities of candidates in time do not integrate to 1.  This is discussed in more detail in \cite{cannon2015}, and so, unlike a true log-odds ratio, the value 0 has no special meaning here.  The ranking statistic is understood to be an arbitrarily normalized ``bigger is better'' scale.\footnote{It differs from the true log-odds ratio by a constant offset which can be estimated, if one wishes, using Wilk's theorem \cite{d543aecb-cd73-36d5-9101-f08a74f8e8c6}, by fitting the distribution of ranking statistic values with the unkonwn offset added to a \(\chi^2\) distribution.}

By drawing from the denominator of the ranking statistic, the GstLAL detection system generates an ensemble of synthetic candidates distributed in the parameter space as the noise process is.  From this ensemble, it predicts the relative frequency with which values of $\ln\mathcal{L}$ are expected to be observed in candidate events produced by noise.  The complementary cumulative distribution function for that density is computed, and from that the measures of significance are derived.

The quantity used to assess the statistical significance of a candidate is the \textbf{\ac{FAP}}.  In this context this is the probability that a signal-free version of the experiment (one in which all candidates are produced by noise) yields at least 1 candidate at or above a given value of $\ln\mathcal{L}$.  A related quantity is the \textbf{\ac{FAR}}.  This is the rate, measured in events per time, at which a signal-free version of the experiment yields candidates at or above a given value of $\ln\mathcal{L}$.  Multiplying the \ac{FAR} by the duration of the experiment, \(T\), gives
\begin{equation}
\left< N \right>
	= \mathrm{FAR} T,
\end{equation}
the number of false positives expected from the noise at or above a given value of $\ln\mathcal{L}$.  The actual number of false positives observed above that threshold should be a Poisson-distributed \ac{RV} with that mean count.

\(\ln\mathcal{L}\), \ac{FAP}, \ac{FAR}, and \(\left< N \right>\) are all related to one another by monotonic 1-to-1 mappings, and are therefore somewhat interchangeable in discussions of significance.

The standard summary of the ensemble of candidates produced by an experiment is a count vs.\@ threshold plot, showing \(\left< N \right>\) vs.\@ \(\ln\mathcal{L}\) threshold predicted by the noise model, with the actual observed count vs.\@ threshold superimposed for comparison.  If the noise model is working well, the two curves should be indistinguishable at low \(\ln\mathcal{L}\) threshold where the event rate is dominated by false positives.  If loud signals are present, signals that have produced statistically-significant candidate events, their presence is revealed by the observed count-above-threshold curve laying above the count predicted by the noise model in the tail.  The plot, therefore, at once, provides a quick visual confirmation of the accuracy of the noise model, and of the presence or absence of signals in the data.

In the tail, where the expected count of false positives due to noise is close to 1, Poisson-distributed random fluctuations in the observed count of false positives become significant, and should not be mistaken for an excess or deficit.  We indicate the scale of these fluctuations with shaded regions showing multiples of \(\sqrt{\left< N \right>}\) above and below \(\left< N \right>\).  \(\sqrt{\left< N \right>}\) is the standard deviation of the distribution, but multiples of \(\pm\sqrt{\left< N \right>}\) do \emph{not} indicate the 68\%, \textit{etc.}, credible intervals for the count.  Because the count must be an integer, generally there is no set of outcomes whose combined probability equals some chosen value, and so a, for example, 68\% interval does not exist, it cannot be shown.  The shaded regions are intended merely to provide a visual clue as to what size of fluctuations is normal, not the specific ranges of values to be expected.


\section{Results}\label{sec:results}

The results of the \ac{S5} and \ac{S6} analysis are summarized in a list of events, which are ranked from most to least significant as determined by GstLAL. In this section, we discuss the most significant event, GW070605, and describe the results of the follow-up investigation that ultimately confirms its non-astrophysical origin. We also measure the sensitivities between \ihope{} and GstLAL for two cases---at the null-result threshold and at thresholds above null-result---in order to provide as complete a picture as possible of the pipelines' performances and how they compare with each other.

\subsection{Triggers list}\label{sec:triggers}

\begin{figure}
\includegraphics[width=\columnwidth]{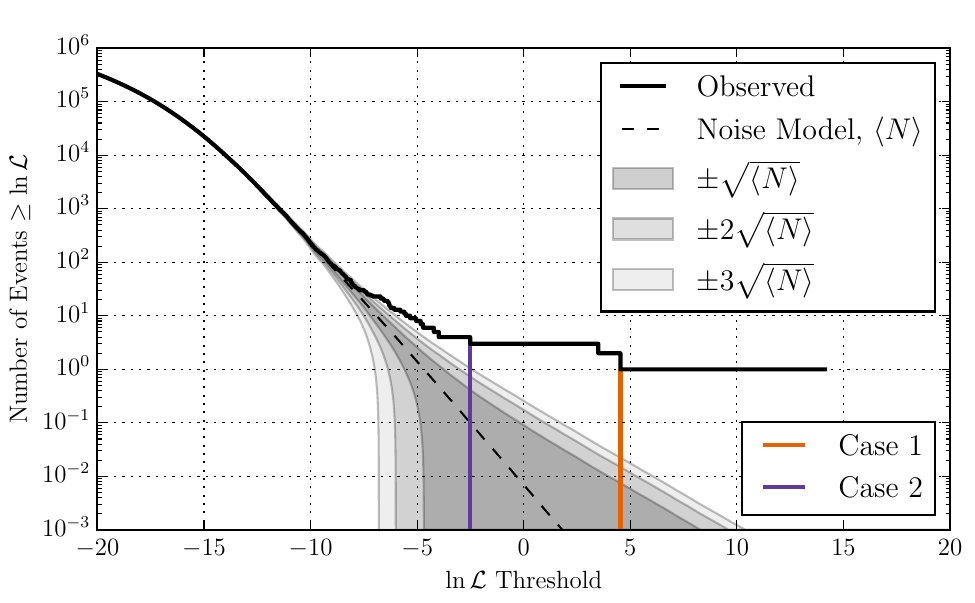}
\caption{Number of events that were recovered with $\ln\mathcal{L}$ ranking statistic values above the threshold. The dashed line represents the predicted number of background noise events (with the shaded regions indicating the Poisson-distributed random fluctuations in multiples of $\pm\sqrt{\langle N\rangle}$), while the solid black line denotes the observed number of events in the analysis. The highest significant event, GW070605, was found with a log likelihood ratio $\ln\mathcal{L}=14.13$. The vertical lines at $\ln\mathcal{L} = 4.54$ (Case 1) and $\ln\mathcal{L} = -2.51$ (Case 2) mark the thresholds for when the top ranked and top three ranked candidates are excluded, respectively. These cases are later used as the null-result $\ln\mathcal{L}$ thresholds in \secref{sec:sensitivity}.}
\label{fig:openbox}
\end{figure}

The main search results are summarized in \figref{fig:openbox}, which plots the number of events whose $\ln\mathcal{L}$ exceed a given threshold. As $\ln\mathcal{L}$ increases, we begin to see the observed counts (solid curve) diverge from the noise model (dashed curve), and at roughly $\ln\mathcal{L}\sim-7$ (where the noise curve is 1), the curves differ by six events. We round up this number to ten and list the top ranked triggers identified by GstLAL in \tabref{table:triggers}, sorted by ascending \ac{FAR}. For each event, its time, the detector(s) that observed it, its \ac{FAR}, \ac{FAP}, $\ln\mathcal{L}$, and the network \ac{SNR} are given. The final five columns describe the chirp mass and component mass and spin parameters of the template that recovered the event. 

In both \figref{fig:openbox} and \tabref{table:triggers}, it is clear that GstLAL identified at least three events with high significance, with a \ac{FAR} of $8.6\times10^{-10}$ per year assigned to the top ranked event. Since its significance well exceeds the threshold of what would be considered a confident \ac{GW} detection, we refer to the event as GW070605. 

\begin{table*}
\centering
\begin{tabular}{|c|c|c|c|c|c|c|c|c|c|c|}
\hline
\multirow{2}{*}{\bf UTC Time} & \multirow{2}{*}{\bf IFO} & \bf \parbox[t]{1.7cm}{\ac{FAR}\\(per year)}& \multirow{2}{*}{\bf FAP} & \multirow{2}{*}{\bf $\ln\mathcal{L}$} & \bf \parbox[t]{1.7cm}{Network\\\ac{SNR}} & \multirow{2}{*}{\bf $\mathcal{M}_c$ ($\Msun$)} & \multirow{2}{*}{\bf $m_1$ ($\Msun$)} & \multirow{2}{*}{\bf $m_2$ ($\Msun$)} & \multirow{2}{*}{\bf $\chi_1$} & \multirow{2}{*}{\bf $\chi_2$} \\
\hline
2007-06-05 18:36:37 & L1 & $8.6\times10^{-10}$ & $2.17\times10^{-9}$ & 14.13 & 13.8 & 1.30 & 1.82 & 1.24 & 0.050 & 0.046 \\
2005-11-23 11:12:14 & L1 & $2.2\times10^{-5}$ & $5.56\times10^{-5}$ & 4.54 & 26.0 & 10.34 & 14.02 & 10.11 & 0.115 & 0.422 \\
2006-05-10 06:46:25 & H1 & $6.5\times10^{-5}$ & $1.63\times10^{-4}$ & 3.50 & 8.5 & 3.14 & 12.21 & 1.32 & -0.341 & 0.041 \\
2006-07-20 17:26:05 & H1 &  $0.040$ & 0.095 & -2.51 & 12.0 & 7.45 & 11.10 & 6.69 & -0.538 & 0.902 \\
2007-01-09 10:01:42 & H1L1 & 0.19 & 0.39 & -3.98 & 9.7 & 2.73 & 12.48 & 1.02 & -0.557 & -0.030 \\
2007-07-08 20:29:28 & H1L1 & 0.24 & 0.46 & -4.22 & 8.1 & 2.63 & 11.53 & 1.01 & 0.074 & -0.005 \\
2010-09-25 14:24:52 & H1L1 & 0.41 & 0.65 & -4.73 & 8.7 & 2.05 & 6.35 & 1.02 & -0.234 & -0.010 \\
2007-03-15 05:21:05 & H1 &  0.46 & 0.69 & -4.83 & 8.6 & 6.59 & 8.15 & 7.04 & -0.990 & -0.828 \\
2009-08-12 09:36:31 & H1L1 & 0.58 & 0.77 & -5.05 & 8.4 & 4.54 & 14.17 & 2.24 & 0.101 & -0.026 \\
2007-08-05 18:15:26 & H1L1 & 0.78 & 0.86 & -5.33 & 8.1 & 1.35 & 2.05 & 1.19 & -0.007 & 0.035 \\
\hline
\end{tabular}
\caption{Table of ten triggers with the lowest \ac{FAR}. The last five columns (chirp mass $\mathcal{M}_c$, component masses $m_1$ and $m_2$, and component spins $\chi_1$ and $\chi_2$) correspond to the parameters of the template that recovered the trigger.}
\label{table:triggers}
\end{table*}

\subsection{GW070605} \label{sec:GW070605}
The story of GW070605 begins when GstLAL reported a candidate event whose end time occurred on June 5, 2007 at 18:37:02 UTC (865103811.84 GPS). Observed in only the L1 detector (H1 was not online at the time) with a \ac{FAR} of $8.6\times10^{-10}$ per year, it is the most significant event detected by GstLAL in this search. The event was recovered by a waveform template having parameters $[m_1, m_2, \chi_1, \chi_2] = [\qty{1.82}{\Msun}, \qty{1.24}{\Msun}, 0.050, 0.046]$, which is consistent with a low-spin \ac{BNS} merger.

\begin{figure}
\includegraphics[width=\columnwidth]{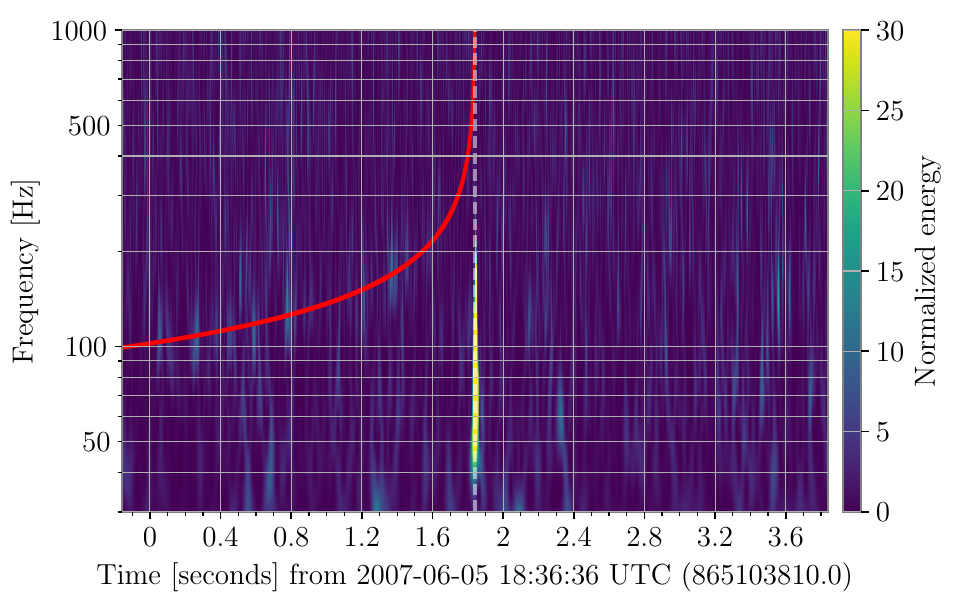} 
\caption{Q-transform spectrogram of L1 strain data around the time of the candidate event GW070605. The red curve denotes the representative waveform of GW070605, with its end time marked by the vertical dashed line. The Q-values used to create the spectrogram are Q$_\mathrm{min} = 4$ and Q$_\mathrm{max} = 64$.}
\label{fig:gw070605_omega}
\end{figure}

GW070605's high significance inspired further follow-up to confirm its origin. As a first test, a timeseries spectrogram computed from the Q-transform \cite{Chatterji_2004,Chatterji2005} was created using GWpy \cite{gwpy} to visually identify stretches of strain data that deviated from a stationary Gaussian noise background. \figref{fig:gw070605_omega} shows the Q-transform spectrogram of L1 strain data around the time of GW070605, revealing the presence of an abrupt, glitch-like spike of excess power across the \qtyrange{30}{200}{\hertz} frequency band. The red curve represents the time evolution of the frequency of the GW070605 template.  The vertical dashed line marks the end time of GW070605.  That dashed line shows that the time of the glitch is extraordinarily close to GW070605's time of merger. Adhering to the philosophy that coincidences should not be trusted \cite{startrekds_S02E05}, the peculiar synchronization of the glitch and coalescence times warranted further investigation.

\begin{figure}
\includegraphics[width=\columnwidth]{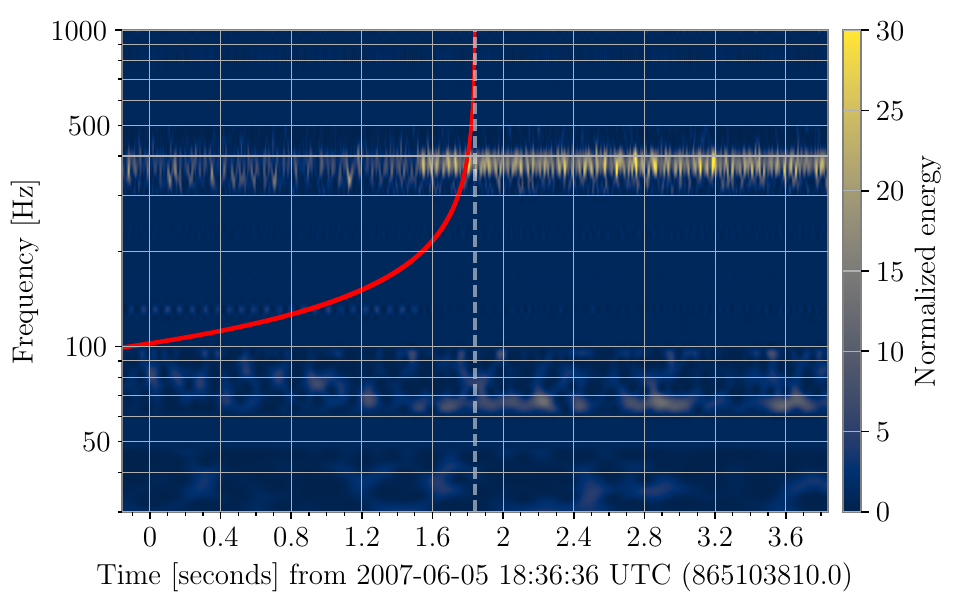} 
\includegraphics[width=\columnwidth]{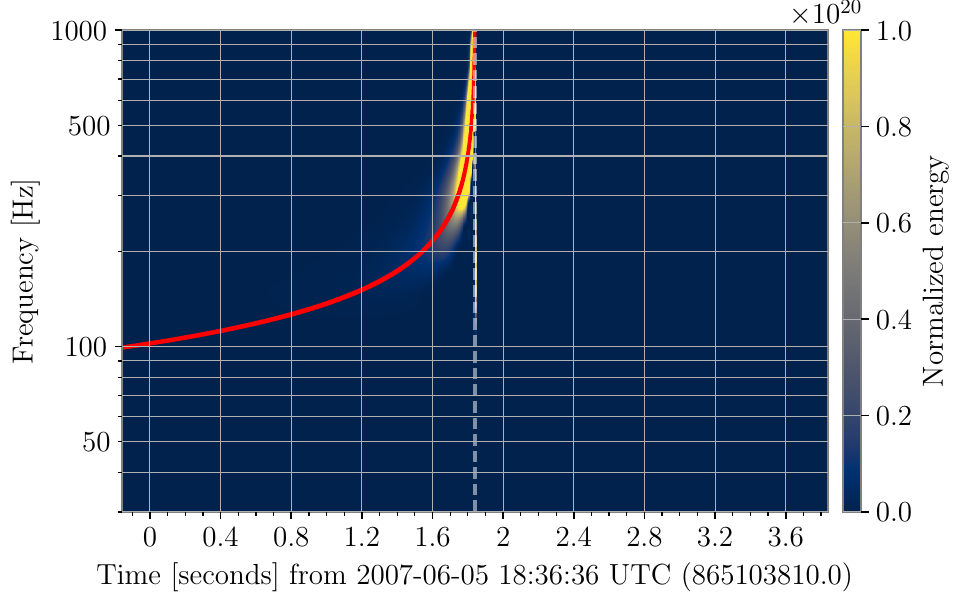} 
\caption{Q-transform spectrogram of the L1 ETMX (top) and ETMY (bottom) coil excitation channels. In both panels, the red curve denotes the representative waveform of GW070605, with its end time marked by the vertical dashed line.}
\label{fig:gw070605_etm}
\end{figure}

One hypothesis we considered was that GW070605 was an improperly constructed hardware injection \cite{Biwer_2017}, with the coincident glitch being the result of the hardware injection team inadvertently terminating the waveform too early, creating a step function at its end.  If this was true, in addition to being malformed, it would be an undocumented hardware injection, since neither \ac{GWOSC} nor \ac{LIGO} list any injections within four hours of GW070605's GPS time. Our attention turned to L1's auxiliary data channels;  in particular, the coil excitation channels of the \acp{ETM} in the X and Y arms, as these coil actuators had been used to launch hardware injections (X arm) and blind hardware injections (Y arm) in S5 and S6 \cite{Berliner2010,Berliner2010_2}.  This investigation required access to proprietary data as the excitation channels are not available through \ac{GWOSC}.  \figref{fig:gw070605_etm} shows the Q-transform spectrograms of the \ac{ETM}X (top  panel) and \ac{ETM}Y (bottom panel) coil excitation channels.  Our smoking gun lies in the bottom panel, where \ac{ETM}Y, which is plotted using a colour scale differing from \ac{ETM}X's by nearly 20 orders of magnitude, shows excess energy that indeed tracks the frequency vs.\ time chirp of GW070605, represented again by the red track.

As only the L1 detector was online at the time, previous analyses could not and were not expected to find GW070605, which might be the reason it was not recorded in injection lists. Since the \ac{ETM}Y coil excitation channel was used for blind injections in S5, we conclude that GW070605 is a malformed, undocumented, single-detector blind injection.

With respect to the glitch, because we see a compact object merger hardware injection in an excitation channel at the time of the event, we are in the unusual position of \emph{knowing for a fact} that GW070605 is a ``real'' compact object merger signal in the strain data of L1.  Given that knowledge, GW070605 teaches some important lessons about glitches and false positives.  The detection system's assessment of the significance of this event was correct.  Second-guessing the detection system's statistical analysis of the data based on the appearance of the strain's time-frequency plot would have been an error.  The signal, which is invisible in the time-frequency plot, \emph{is} present in the data, and the candidate is \emph{not} a false positive created by a glitch.   The presence of the glitch feature in the time-frequency plot, and the absence of a visible chirp track, are uninformative.  Vetoing a possible signal because it possess these characteristics would have been unwise.

\begin{figure}
\includegraphics[width=\columnwidth]{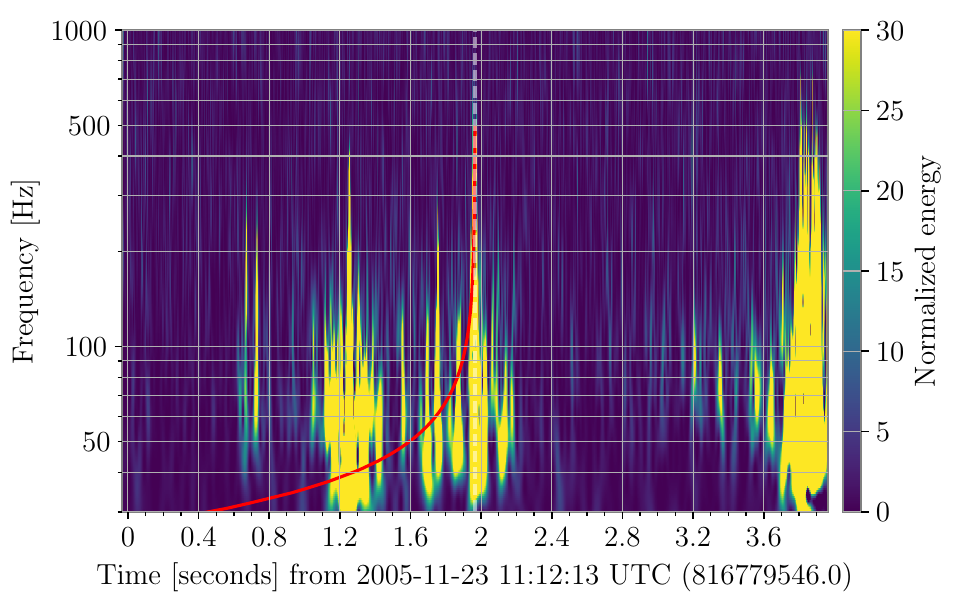}
\includegraphics[width=\columnwidth]{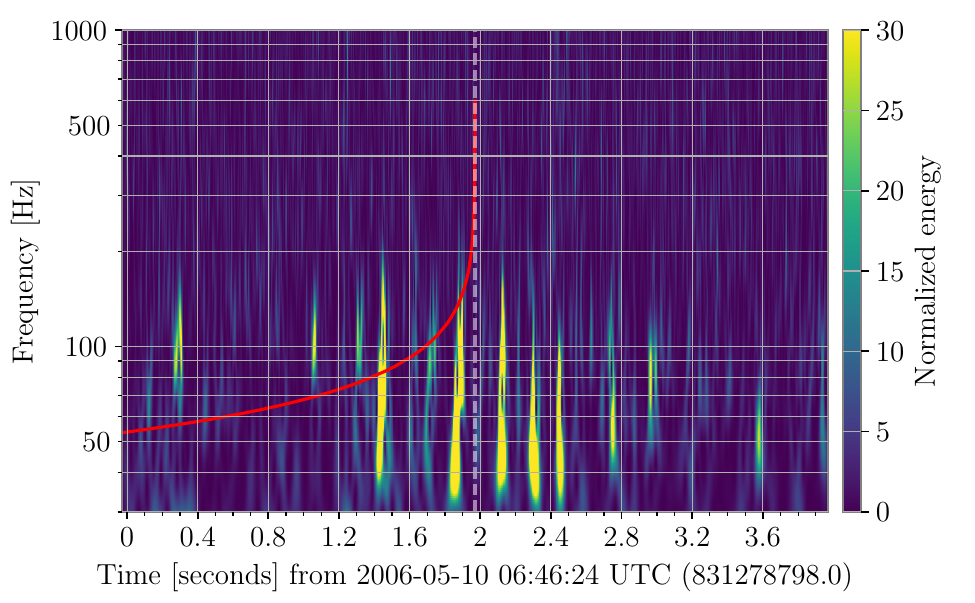}
\caption{Q-transform spectrograms of the 2nd and 3rd top ranked candidates in \tabref{table:triggers}. In both panels, the red curve denotes the template waveform that recovered the candidate, with its end time marked by the vertical dashed line.}
\label{fig:omega_othertriggers}
\end{figure}

This brings us to the next two candidates in the list.  The detection algorithm has also determined them to be statistically significant. As shown in \figref{fig:omega_othertriggers}, there are also glitch features visible in their time-frequency plots, and those plots also lack obvious chirp tracks.  They seem to not be the result of hardware injections because the excitation channels recorded for those times are 0, which is normal when there are no injected signals \cite{Evans2015}.  We conclude nothing more about these candidates at this time.  Those plots, however, also remind the reader of the existence of (many) other glitches in the data that are \emph{not} associated with high significance candidates. 

\subsection{Sensitivity}\label{sec:sensitivity}

The original objective of this endeavour was to compare the sensitivity of a modern detection algorithm to the sensitivity of the detection algorithms of the time.  We expected to do this by extracting a null-result rate upper bound from our search and comparing it to the previously reported rate bound.  The ratio of the two is the ratio of the signal detection rates of the two algorithms at the null-result threshold.  We did not obtain a null result, since the detection algorithm identified at least one genuine (so to speak) signal, and this has upset the apple cart.

Identifying a previously undocumented ``signal'' in \ac{S5} and \ac{S6} data that has gone unreported for 18 years is a dramatic demonstration of the capabilities of a modern detection algorithm like GstLAL.  However, concretely this is little more than an anecdotal example of the performance improvements, and we still wish to rigorously quantify the difference.


\subsubsection{At Null-result Threshold}\label{sec:nullresult}

The null-result bound on the rate of signals is obtained by setting a threshold at the ranking statistic value of the highest-ranked event observed by the search, and measuring the spacetime volume, \(VT\), within which simulated signals can be recovered with ranking statistic values above that threshold.  Since 0 candidates were observed above that threshold, the number of signals inside that \(VT\) is \(< 1\) with some confidence, or the upper bound on the rate of signals per volume per time is 
\begin{equation}
R_{90}
	= \frac{C}{VT},
	\label{eq:R_90_Bayesian}
\end{equation}
where the proportionality constant $C$ depends on the desired confidence and some statistical assumptions.  In past searches, three forms of this relationship have been documented \cite{Biswas_2009,bradycreightonwiseman2004}, where $C=[2.303, 3.272, 3.890]$ for the frequentist, transitional, and Bayesian cases, respectively.

Instead of removing one or more non-false-positive candidates from our result and reporting a rate upper bound derived from those that remain, we chose to convert the previous \ihope{} result back to a \(VT\) accessible above threshold. 
To compute a \(VT\), we made two choices of threshold: 
\begin{enumerate}
\item The \(\ln\mathcal{L}\) ranking statistic value of the loudest event that remains after removing GW070605, the one candidate we know with certainty is not a false positive, and
\item The \(\ln\mathcal{L}\) of the loudest event remaining after excluding the three high-significance outliers.  We know there to be at least one undocumented hardware injection in these data, and ignoring the other two statistically significant candidates is an easier path forward than trying to diagnose their origin.  The loudest event that remains is still a moderate, ``\(2 \sigma\)'', outlier, similar in significance to outliers observed in previous searches that were also treated as null results, for example \cite{Abadie_2011}.
\end{enumerate}

Several astrophysically motivated populations of mergers were simulated and injected into the strain data.  Using the waveform model IMRPhenomD, the injections span the full parameter space of detectable signals, which encompass \ac{BNS}, \ac{BBH}, and \ac{NSBH} sources:
\begin{enumerate}
\item{\textbf{\Ac{BNS}}
	\begin{itemize}
	\item{Set A (``narrow \ac{BNS}"): Gaussian-distributed mass, with mean $\mu = \qty{1.33}{\Msun}$ and standard deviation $\sigma^2 = 0.0081$, with limits $\qty{0.8}{\Msun} \le m_{1,2} \le \qty{2.3}{\Msun}$. Uniformly distributed spin, with limits $|\chi_{1,2}| \le 0.05$. This injection set matches the distributions described in \cite{Ozel_2016}.}
	\item{Set B (``broad \ac{BNS}"): Uniformly distributed mass, with limits $\qty{1}{\Msun} \le m_{1,2} \le \qty{3}{\Msun}$. Uniformly distributed spin, with limits $|\chi_{1,2}| \le 0.05$. This set covers a larger mass parameter space for \ac{BNS} sources.}
	\end{itemize}}
\item{\textbf{\Ac{BBH}}
	\begin{itemize}
	\item{Set C (``uniform \ac{BBH}"): Uniformly distributed mass, with limits $\qty{5}{\Msun} \le m_{1,2} \le \qty{50}{\Msun}$. Uniformly distributed spin, with limits $|\chi_{1,2}| \le 0.99$.}
	\item{Set D (``logarithmic \ac{BBH}"): Logarithmically distributed mass, $\qty{1}{\Msun} \le m_{1,2} \le \qty{99}{\Msun}$. Uniformly distributed spin, with limits $|\chi_{1,2}| \le 0.99$. This set also includes \ac{IMBH} sources.}
	\end{itemize}}
\item{\textbf{\Ac{NSBH}}
	\begin{itemize}
	\item{Set E: Uniformly distributed mass, with limits $\qty{3}{\Msun} \le m_{1} \le \qty{7}{\Msun}$, $\qty{1.3}{\Msun} \le m_{2} \le \qty{1.4}{\Msun}$. Uniformly distributed spin, with limits $|\chi_{1}| \le 0.99$, $|\chi_{2}| \le 0.05$. This set was created in order to match the NSBH mass distributions used in the previous \ac{S5} and \ac{S6} analyses \cite{Abadie_2010,Abadie_2012}.}
	\end{itemize}}
\end{enumerate}
\tabref{table:ihope_mass_distributions} reproduces \ihope{}'s rates upper limits as first reported in \cite{Abadie_2012}, as well as the mass distributions they used to define \ac{BBH}, \ac{NSBH}, and \ac{BNS} systems, which are narrower than the parameter space covered by our abovementioned injection sets. As a result, of all the injections in Sets A--E, only those that satisfy the mass distributions listed in \tabref{table:ihope_mass_distributions} are used in our results. 
\begin{table}
\centering
\begin{tabular}{|c | c | c |c |}
\hline
 & $R_{90}$ (Mpc$^{-3}$ yr$^{-1}$) & $m_1$ ($\Msun$) & $m_2$ ($\Msun$) \\
\hline
BBH & $7.4\times10^{-6}$ & $5 \pm 1$ & $5 \pm 1$ \\
NSBH & $3.6\times10^{-5}$ & $1.35 \pm 0.05 $ & $5 \pm 1 $ \\
BNS & $1.3\times10^{-4}$ & $1.35 \pm 0.05$ & $1.35 \pm 0.05$ \\
\hline
\end{tabular}
\label{table:ihope_mass_distributions}
\caption{\ihope{}'s 90\% confidence upper bound on the rates of signals, $R_{90}$ (non-spinning for BNS, and spinning for NSBH and BBH), and mass distributions of each system \cite{Abadie_2012}. Although the injections used by GstLAL span the full parameter space of detectable signals, only injections that match the same mass distributions used by \ihope{} are included to calculate GstLAL's sensitive time-volume curves in \figref{fig:vt_vs_lnL}.}
\end{table}

\begin{figure}
\includegraphics[width=\columnwidth]{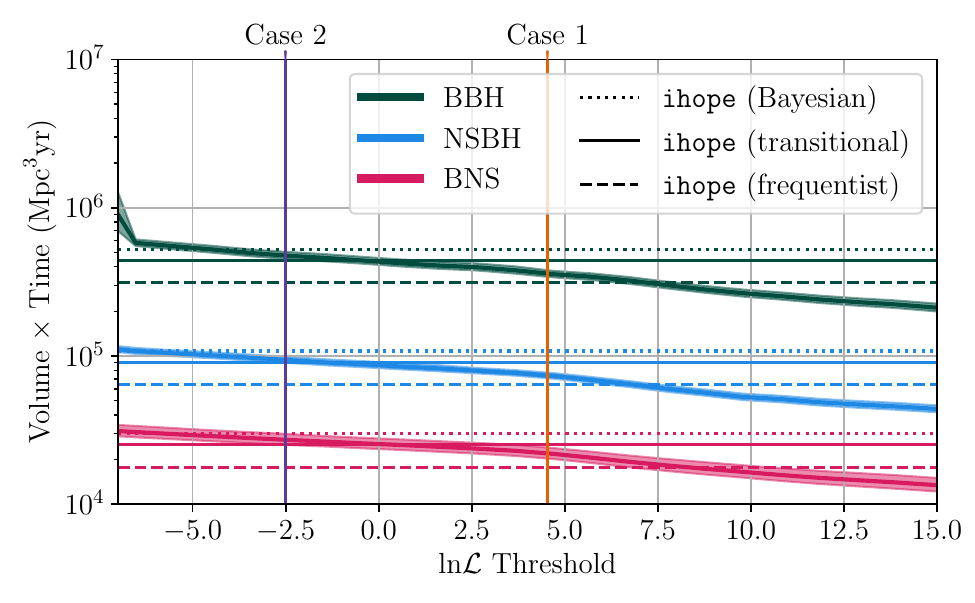}
\caption{The sensitive time-volume $VT$, within which signals drawn from the injection population are recovered with ranking statistic values above the $\ln\mathcal{L}$ threshold. The solid curves represent GstLAL's $VT$, where the mass distributions of each system (BBH, NSBH, and BNS) are specified in \tabref{table:ihope_mass_distributions}; the shaded regions indicate the 68\% confidence interval. The horizontal lines represent \ihope{}'s upper limit $VT$ at its null-result threshold for each system, which are estimated from the $R_{90}$ rates listed in \tabref{table:ihope_mass_distributions} and \eqref{eq:R_90_Bayesian}, where the choice of proportionality constant $C$ is differentiated by line style. The vertical solid lines mark two choices of threshold: the loudest event that remains when GW070605 is excluded (Case 1), and the loudest event that remains after excluding the top three most significant outliers (Case 2).}
\label{fig:vt_vs_lnL}
\end{figure}

As a function of ranking statistic threshold, the \(VT\) visible to the GstLAL search 
is shown in \figref{fig:vt_vs_lnL}. 
The shaded area indicates the uncertainty in \(VT\).  Two processes contribute to the measurement uncertainty:
\begin{enumerate}
\item Uncertainty in the calibration of the detectors, which leads to an uncertainty in the relationship between the physical distance to a simulated source and the amplitude with which its \ac{GW} strain should appear in the data during the test.  Strain amplitude is inversely proportional to distance, and volume depends cubically on distance so \(VT\) is particularly sensitive to strain calibration uncertainty.
\item The binomially-distributed noise in the count of successfully recovered simulations arising from the finite number of them.  Normally, this is a sub-dominant effect by construction --- enough injections are done to ensure the measurement uncertainty is dominated by calibration uncertainty.  Jaggedness in the \(VT\) vs.\@ threshold curve tends to be caused by this noise source, not genuine changes in sensitivity at certain threshold values.
\end{enumerate}
For the purpose of comparing to the sensitivity of a previous search conducted with the same data, the calibration uncertainty is irrelevant as it affects both search results identically:  the effect of a miscalibrated detector factors out of the ratio of the two searches' \(VT\).  Only sampling noise due to the finite number of simulations used in both tests leads to errors in the measured \(VT\) ratio.  The two contributions were not reported separately in the previous search results, so we can only isolate the sampling noise contribution by guessing, which we do not attempt.  We have set the calibration uncertainty to 0 when computing our own \(VT\) from the found and missed injection lists.  This leads to a \(VT\) estimate that appears to be much more precisely measured than what was previously reported, and that difference should be ignored.

Using \eqref{eq:R_90_Bayesian} and \ihope{}'s reported $R_{90}$ values (which are reproduced in \tabref{table:ihope_mass_distributions}), we estimate \ihope{}'s \(VT\) upper limit for each system. Because the value of the proportionality constant used to compute \ihope{}'s rate upper limits in \cite{Abadie_2012} is not stated, we consider all available choices, yielding an estimated \(VT\) upper limit range of $3.1 - 5.3\times10^{5}$ Mpc$^{-3}$ yr$^{-1}$ (\ac{BBH}), $0.6 - 1.1\times 10^{5}$ Mpc$^{-3}$ yr$^{-1}$ (\ac{NSBH}), and $1.8 - 3.0\times10^{-4}$ Mpc$^{-3}$ yr$^{-1}$ (\ac{BNS}); these values are the horizontal lines in \figref{fig:vt_vs_lnL}, where they plotted alongside GstLAL's \(VT\) at the two choices of $\ln\mathcal{L}$ threshold.

The conclusion is that GstLAL's \(VT\) at null-result threshold lies between the ranges of the previous \ihope{} pipeline's \(VT\).  This conclusion seems paradoxical given that the modern pipeline was able to search nearly twice as much data and succeeded in finding a ``signal'' (undocumented hardware injection) that was missed by the older algorithm.  How do we reconcile the conclusion that the two algorithms' sensitivities are more or less identical, with the obvious difference in their outcomes?

The null-result sensitivity comparison is not conducted at an operating point at which signals would ever be detected.  Indeed, the measurement is by construction performed at the operating point where signals just exactly become indistinguishable from background noise.  For the purpose of deriving upper bounds on compact object merger rates from non-detections the two pipelines are evidently equivalent --- for both, signals become indistinguishable from noise at about the same physical distance from Earth --- but given the different outcomes, we have reason to believe they are not equivalent in their sensitivity to detectable signals:  \(VT\) might be different at operating points corresponding to useful signal detection thresholds.


\subsubsection{At Thresholds Above Null-result}\label{sec:nonnullresult}

Unfortunately, one feature the \ihope{} pipeline lacked was the ability to measure its sensitivity as a function of detection threshold.  A specific threshold had to be chosen, and \(VT\) would be measured only at that threshold.  No plots showing \(VT\) vs.\@ threshold like those shown in \figref{fig:vt_vs_lnL} are available for \ihope{}.

We have one small opportunity to estimate \ihope{}'s sensitivity at a different threshold.  During \ac{S6}, \ihope{} identified a \ac{GW} candidate event, GW100916, in the H1 and L1 detectors and assigned it a combined \ac{FAR} of 1 per 7000 years (or $1.43\times10^{-4}$ per year) \cite{GW100916, Abadie_2012}. Ultimately, that candidate was confirmed to be a blind hardware injection, not an astrophysical signal.  The parameters of that hardware injection are known, and so for our purposes, this provides one single data point mapping distance-to-source to \ac{FAR} for \ihope{}.  The signal was injected into H1, L1, and V1 data 
at a distance of \qty{9.74}{\mega\parsec} \cite{blindinjection,LIGOScientific:2013yzb}.  We address the following questions: at what \ac{FAR} does GstLAL recover the hardware injection, and at what distance are signals recovered by GstLAL with \ihope{}'s \ac{FAR} of $1.43\times10^{-4}$ per year?

As described \secref{sec:pipelines}, data containing (known) hardware injections were omitted from our analysis, including GW100916. For this investigation, we performed a separate analysis that includes GW100916 and excludes all other injections and CAT1 vetoes. The GstLAL pipeline identified the GW100916 with a much higher significance than \ihope{}, reporting a \ac{FAR} of $2.72\times 10^{-23}$ per year. \tabref{table:blindinjection} summarizes the search results of both \ihope{} and GstLAL.

\begin{table*}
\centering
\begin{tabular}{| c | c | c | c |}
\hline
& \multirow{2}{*}{\bf Injected parameters} & \multicolumn{2}{c|}{\bf Detected parameters} \\
&  & \multicolumn{1}{c}{\bf\ihope{}} & \bf GstLAL \\
\hline
Time & Sep 16, 2010 06:42:23 UTC\footnote{Geocentric time} & Sep 16, 2010 06:42:23 UTC & Sep 16, 2010 06:42:23 UTC \\
Network \ac{SNR} & 18 & 15\footnote{This value is derived from a non-public publication \cite{GW100916_supp}---the highest matched-filter SNRs quoted in \cite{Abadie_2012} refer to non-coincident triggers} & 18.6 \\
Chirp mass & $\qty{4.95}{M_\odot}$ & $\qty{3.48}{M_\odot}$ to $\qty{7.40}{M_\odot}\footnote{Mass range corresponding to the mass bin that recovered GW100916}$ & $\qty{4.99}{M_\odot}$\footnote{Chirp mass of the template waveform that recovered GW100916}\\
\ac{FAR} & | & $1.43\times10^{-4}$ per year & $2.72\times 10^{-23}$ per year \\
\hline
\end{tabular}
\caption{Search results of the blind hardware injection GW100916 for the \ihope{} and GstLAL detection pipelines. The blind injection parameters are documented in \cite{blindinjection} and are listed here for reference only; they are neither necessary nor used in our analysis.}
\label{table:blindinjection}
\end{table*}

\begin{figure}
\includegraphics[width=\columnwidth]{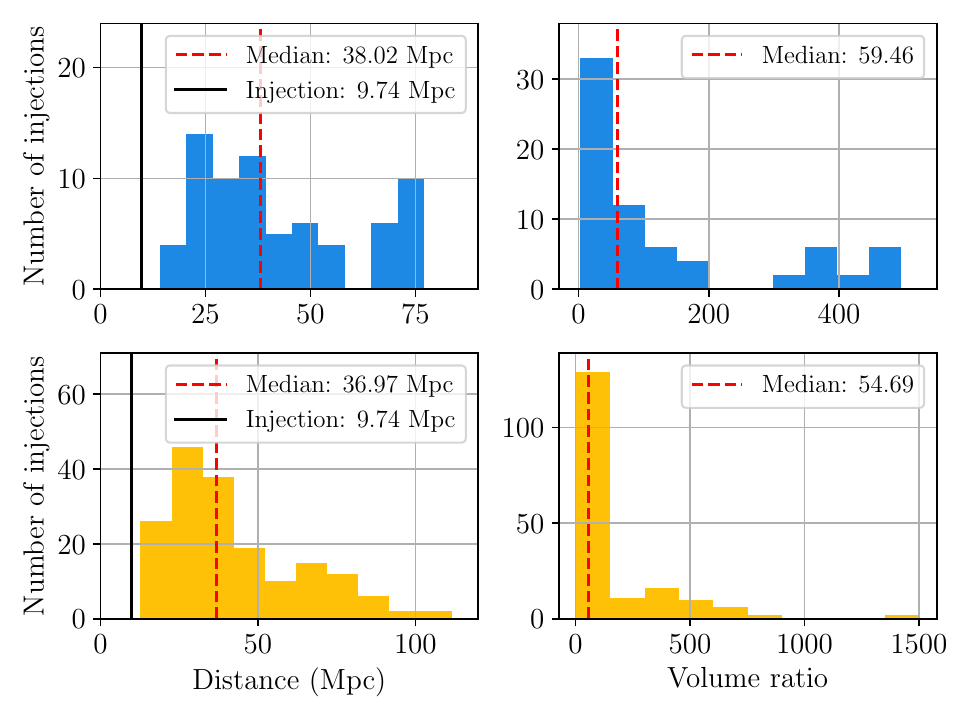}
\caption{Distribution of distances and volume ratios of the injections whose chirp masses fall within the limits of the \ihope{} mass bin that recovered the blind injection GW100916. The median value of each distribution is marked by a vertical red dashed line. The top panels (with blue histograms) show the distributions of 71 injections whose \ac{FAR} values are in the range $\qtyrange{4e-12}{5e-12}{\hertz}$, while the bottom panels (with yellow histograms) show the distributions of 176 injections with an extended \ac{FAR} range of $\qtyrange{3e-12}{6e-12}{\hertz}$.}
\label{fig:nonnullhist}
\end{figure}

The \ihope{} pipeline identified GW100916 with a combined \ac{FAR} of $1.43\times10^{-4}$ per year (or $\qty{4.53e-12}{\hertz}$) in the mass bin $\qty{3.48}{M_\odot} \le \mathcal{M}_c \le \qty{7.40}{M_\odot}$. To determine the average distance to which GstLAL is sensitive at the same \ac{FAR} value, we isolate the injections (from the injection run outlined in \secref{sec:sensitivity}) belonging to the same mass bin that were detected by GstLAL within two different \ac{FAR} ranges: a narrower range of $\qtyrange{4e-12}{5e-12}{\hertz}$ and an extended range of $\qtyrange{3e-12}{6e-12}{\hertz}$. To compare the sensitivity, we estimated the volume ratio by dividing the cube of the distances of the selected injections by the cube of the distance of GW100916. \figref{fig:nonnullhist} shows the distributions of the distances and volume ratios; the top panels show the case for the narrow \ac{FAR} range, for which there are 71 injections, and the bottom panels show the case for the extended \ac{FAR} range, for which the number of injections has increased to 176.

The injections GstLAL recovered that fall within the target \ac{FAR} intervals were found to have been injected over a wide range of distances.  The median distance at which GstLAL recovers injections within the target \ac{FAR} intervals was about \qty{37}{\mega\parsec}.  Compared to GW100916's injected distance of \qty{9.74}{\mega\parsec}, we find the median volume ratio to be approximately $55$ times larger, \textit{i.e.} the GstLAL pipeline recovers 55 times more signals at a \ac{FAR} threshold of 1 per 7000 years than the \ihope{} pipeline.

This appears to offer a story that resolves for us the paradoxical observation that GstLAL recovered a previously unidentified ``signal'' (hardware injection) while apparently having the same \(VT\) as \ihope{} at the null-result threshold.  Although the GstLAL detection pipeline does not differ from the earlier \ihope{} pipeline's sensitivity in the point at which signals and noise become indistinguishable from each other, the transition from high significance signals to unidentifiable signals occurs more sharply in the modern pipeline, closer to that horizon, so that what would be marginal signals in the older pipeline are confident detections in the new.  We find that at a low \ac{FAR} threshold of 1 per 7000 years, the GstLAL pipeline recovers 55 times more signals than \ihope{}---for every 55 signals recovered by GstLAL at or above that significance, all but 1 of them will be recovered at lower significance by \ihope{} or missed altogether, and evidently, GW070605 was an example of one.


\section{Conclusion}\label{sec:conclusion}

Not unlike reopening a cold case with fresh forensics, we have shown the merit of re-analyzing initial \ac{LIGO} data using a modern pipeline.  What was initially an exercise to compare pipeline sensitivities took a detour when GstLAL identified a new single-detector event, GW070605, that was not reported by previous search pipelines. Further investigation into \ac{LIGO}'s auxiliary channels revealed GW070605 to be an undocumented blind hardware injection, and because search algorithms at the time required coincidence between two or more instruments, it is unsurprising that GW070605 went undetected by previous pipelines. After we reported our results to the LVK Collaboration, we are pleased to say that GW070605 is now included in GWOSC's list of S5 injections \cite{gwosc_GW070605}. We direct readers to \url{https://gwosc.org/s5hwcbc/} for the current list of S5 hardware injections.

To compare detector pipeline sensitivities, sensitive time-volume \(VT\) was used as a metric, where we performed an injection campaign (with simulated signals that spanned the full detectable parameter space) to compute GstLAL's \(VT\). The availability of published results from \ihope{} limited our comparison to specific thresholds: null-result and above null-result. While \(VT\)s are comparable between pipelines for the former, at the above null-result threshold, we determined that GstLAL recovers approximately 55 times more signals than the \ihope{} pipeline for signals recovered at a \ac{FAR} threshold of 1 per 7000 years (\textit{i.e.} the \ac{FAR} reported by \ihope{} for the blind injection, GW100916). We thus conclude that there is marked improvement in sensitivity at the above null-result threshold, where GstLAL recovers more signals at greater significance.

Lastly, we conducted a brief test to compare GstLAL's sensitivity with and without Virgo data, whose first three science runs overlapped with \ac{LIGO}'s \ac{S5} and \ac{S6}. Our results are reported in Appendix \ref{app:virgo}, where we conclude that Virgo's contribution improves \(VT\) by up to two times in \ac{S6}/VSR2. We note that ours is an incomplete test, as only two months of data were analyzed, and a complete sensitivity study would comprise a full re-analysis of the data taken during \ac{S5}, \ac{S6}, and VSR1-3.

\acknowledgments
The authors would like to acknowledge Carl-Johan Haster as the catalyst of this study, as well as Aaron Zimmerman for his helpful insights. We would also like to thank the Virgo Collaboration for allowing use of VSR1 and VSR2 data, and the authors of \cite{GW100916} for their work on GW100916. HF acknowledges Stuart Anderson, Juan Barayoga, Evan Goetz, Dan Kozak, Cody Messick, and Rick Savage for their expertise and technical support. HF is supported by the NSERC Alliance International Collaboration program. LT acknowledges NASA 80NSSC23M0104 and the Nevada Center for Astrophysics for support. 

The authors wish to acknowledge that at the time of publication, a handful of citations are not available to the public, but have been included in anticipation that they will one day be made so. The conclusions in this paper do not depend on these citations.

This research has made use of data or software obtained from the Gravitational Wave Open Science Center (gwosc.org), a service of the \ac{LIGO} Scientific Collaboration, the Virgo Collaboration, and KAGRA. GstLAL was developed with support from the National Science and Engineering Research Council of Canada (NSERC) and the Canadian Institute for Advanced Research (CIAR).  The authors are grateful for computational resources provided by the LIGO Laboratory at Caltech, LHO, and LLO, which are supported by the National Science Foundation Grant No. PHY-2309200.  Spectrograms were made using the GWpy software (version 3.0.11) \cite{gwpy} and LALSuite \cite{lalsuite}. Colourblind accessible palettes for figures, courtesy of David Nichols \cite{Nichols}. Finally, this material is based upon work supported by NSF's \ac{LIGO} Laboratory which is a major facility fully funded by the National Science Foundation, as well as the Science and Technology Facilities Council (STFC) of the United Kingdom, the Max-Planck-Society (MPS), and the State of Niedersachsen/Germany for support of the construction of Advanced \ac{LIGO} and construction and operation of the GEO600 detector. Additional support for Advanced \ac{LIGO} was provided by the Australian Research Council. Virgo is funded, through the European Gravitational Observatory (EGO), by the French Centre National de Recherche Scientifique (CNRS), the Italian Istituto Nazionale di Fisica Nucleare (INFN) and the Dutch Nikhef, with contributions by institutions from Belgium, Germany, Greece, Hungary, Ireland, Japan, Monaco, Poland, Portugal, Spain. KAGRA is supported by Ministry of Education, Culture, Sports, Science and Technology (MEXT), Japan Society for the Promotion of Science (JSPS) in Japan; National Research Foundation (NRF) and Ministry of Science and ICT (MSIT) in Korea; Academia Sinica (AS) and National Science and Technology Council (NSTC) in Taiwan.

\appendix

\section{Sensitivity including Virgo data} \label{app:virgo}

A shorter analysis was performed to investigate whether or not the inclusion of Virgo data affects GstLAL's sensitivity. Taking into consideration that the noise behaviour of the \ac{LIGO} and Virgo detectors varied between science runs, we analyzed one set comprising 30 days of data from \ac{S5} (Jun 24, 2007 16:06:37 UTC to Jul 24, 2007 16:06:37 UTC) and a second set comprising 30 days of data from \ac{S6} (Jul 9, 2010 21:00:00 UTC to Aug 6, 2010 21:00:00 UTC). These analysis times were chosen in order to overlap with Virgo's first and second science runs, VSR1 and VSR2; we note that at the time of this publication, VSR1 and VSR2 data are not publicly available. The breakdown of analysis time for each detector network configuration--with and without Virgo---is shown in \tabref{table:time_virgo}. The inclusion of Virgo data adds 1.95 days of analysis time to S5/VSR1 and 7.86 days of analysis time to S6/VSR2, corresponding to an increase of 7\% and 37\% to their respective sets. The total coincidence analysis time is also extended by 16\% (S5/VSR1) and 32\% (S6/VSR2).

\begin{table}
\centering
\begin{tabular}{| c | c | c | c |}
\hline
{\bf Set} & {\bf Instruments} & \bf H1L1 & \bf H1L1V1 \\
\hline
\multirow{5}{*}{\ac{S5}/VSR1} & H1L1V1 & | & 15.29 days \\
 & H1L1 & 19.61 days & 4.31 days \\
 & H1V1 & | & 4.04 days \\
 & L1V1 & | & 2.02 days \\
 & H1 & 5.02 days & 0.98 days \\
 & L1 & 2.53 days & 0.51 days  \\
 & V1 & | & 1.96 days\\
 \cline{2-4}
& \bf{Total} & 27.16 days & 29.11 days \\
\hline
\multirow{5}{*}{\ac{S6}/VSR2} & H1L1V1 & | & 6.67 days \\
 & H1L1 & 7.47 days & 0.80 days \\
 & H1V1 & | & 4.45 days \\
 & L1V1 & | & 7.70 days \\
 & H1 & 4.97 days & 0.52 days \\
 & L1 & 8.72 days & 1.02 days  \\
 & V1 & | & 7.86 days \\
\cline{2-4}
& \bf{Total} & 21.16 days & 29.02 days \\
\hline
\end{tabular}
\caption{Analysis time for each detector network configuration, with and without Virgo. The start and end times for each set are as follows: Jun 24, 2007 16:06:37 UTC to Jul 24, 2007 16:06:37 UTC (S5/VSR1) and Jul 9, 2010 21:00:00 UTC to Aug 6, 2010 21:00:00 UTC (S6/VSR2).}
\label{table:time_virgo}
\end{table}

\begin{figure}
\includegraphics[width=\columnwidth]{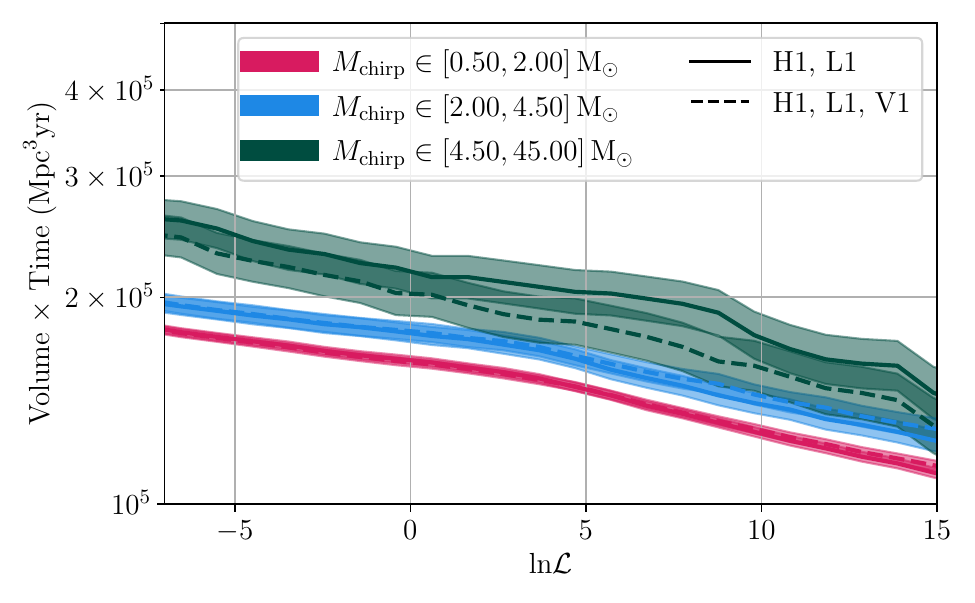}
\includegraphics[width=\columnwidth]{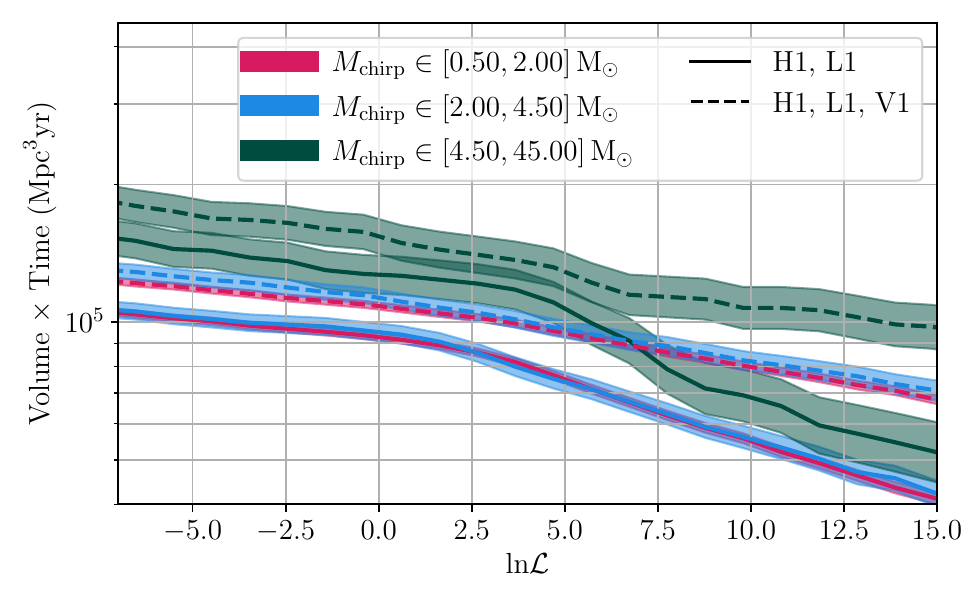}
\caption{Sensitive time-volume $VT$ as a function of the ranking statistic $\ln\mathcal{L}$ for \ac{S5} (top) and \ac{S6} (bottom). The solid curves denote the $VT$ determined from analyzing only Hanford (H1) and Livingston (L1) detector data, and the dashed curves represent the $VT$ determined from analyzing data from a three-detector network---H1, L1, and Virgo (V1). The shaded regions indicate the 68\% confidence interval. The colour of the curves and shaded regions correspond approximately to \ac{BNS} (pink), \ac{NSBH} (blue), and \ac{BBH} (green).}
\label{fig:VT_virgo}
\end{figure}

As before, we compare the sensitivity by computing $VT$s, utilizing the same injection sets described in \secref{sec:results} for the injection campaign. Results are shown in \figref{fig:VT_virgo}, where \(VT\) is plotted as a function of the ranking statistic $\ln\mathcal{L}$ in the \ac{S5} set (top panel) and \ac{S6} set (bottom panel). The injections are categorized into the usual \ac{BNS}, \ac{NSBH}, and \ac{BBH} source types; however, since we are not bound by the mass ranges used by \ihope{} in this analysis (as was the case for \figref{fig:vt_vs_lnL}), we follow the more conventional chirp mass ranges here.  In the \ac{S5} set, where the with- and without-Virgo shaded regions overlap the $VT$ are statistically indistinguishable. The test indicates no improvement when Virgo data is included in the \ac{S5} set. This is unsurprising given the sensitivity of Virgo during its first science run; we direct the readers to Figure 2 of \cite{Abadie_2012} which illustrates the sensitivities of each detector during each science run. However, by the time of VSR2, its second science run, Virgo's sensitivity has more than doubled, an improvement that is also reflected in the bottom panel of \figref{fig:VT_virgo}. 

\bibliography{article}
\end{document}